\documentclass[fleqn,usenatbib]{rasti}

\usepackage{newtxtext,newtxmath}

\usepackage[T1]{fontenc}

\DeclareRobustCommand{\VAN}[3]{#2}
\let\VANthebibliography\thebibliography
\def\thebibliography{\DeclareRobustCommand{\VAN}[3]{##3}\VANthebibliography}

\usepackage{graphicx}	
\usepackage{amsmath}	

\title[Photon Interval Statistics]{Photon Interval Statistics Measure Rapid Variability}

\author[J. I. Katz and M. Nowak]{
J. I. Katz,$^{1}$\thanks{E-mail: katz@wuphys.wustl.edu}
M. Nowak,$^{1}$
\\
$^{1}$Department of Physics and McDonnell Center for the Space Sciences
Washington University, 1 Brookings Dr., St. Louis, Mo. 63130 USA
}

\date{Accepted XXX. Received YYY; in original form ZZZ}

\pubyear{2025}

\begin{document}
\label{firstpage}
\pagerange{\pageref{firstpage}--\pageref{lastpage}}
\maketitle

\begin{abstract}
	Modern X-ray and gamma-ray observatories time-tag detected photons.
	The distribution of intervals between successive photons may reveal
	variations of the flux on time scales too short for direct flux
	measurement of the mean count rate, provided a sufficient number of
	photons have been detected cumulatively.  We demonstrate this with
	synthetic data and apply to RXTE data from Cyg X-1.
\end{abstract}




\section{Introduction}
The variability of an astronomical object is a strong constraint on its
internal dynamics.  Rapid variability indicates a combination of small
dimensions and high velocities.  It is therefore desirable to measure
variability on as short a time scale as possible.  If the flux varies on
some time scale then the statistics of photon detections will be
non-stationary on that time scale, but if it does not vary the statistics
will be Poissonian (shot-noise).

Modern detectors of near-infrared, visible, ultraviolet, X-ray and gamma-ray
radiation count individual photons.  The flux of an astronomical source is
calculated from the counts of a photon-counting detector in a time interval,
provided the detector's characteristics are known.  Measuring the flux with
fractional accuracy $f$ and signal-to-noise ratio $S/N$ requires detecting
about $(S/N)^2/f^2$ photons, assuming no systematic error, and requires
averaging over a time
\begin{equation}
	\label{t}
	t \gtrapprox {(S/N)^2 \over f^2 F},
\end{equation}
where $F$ is the mean count rate; the number of counted photons
\begin{equation}
	\label{N}
	N_{phot} \gtrapprox {(S/N)^2 \over f^2}.
\end{equation}
This limits direct detection of very rapid, or small amplitude, variability
from the time dependence $F(t)$ of the flux.

An alternative approach measures the statistics of the intervals between
successive detected photons.  The limits of Eqs.~\ref{t} and \ref{N} still
apply, but now refer to the length and total number of photons detected in
an entire dataset or observation.  The price paid is that while rapid or
small amplitude variability may be inferred, it is impossible to determine
when within the observation the variation occurred.  This may be less
important than the existence or absence of variability.

Often the duration of an observation is orders of magnitude longer than
interesting time scales of variability, so that many more photons are
detected than the minimum $N_{phot}$ of Eq.~\ref{N}.  This is particularly
likely to occur in observations of compact objects such as accretional X-ray
sources, whose characteristic time scales may be ${\cal O}
(30\,\mu\text{s})$, those of infall onto neutron stars or stellar-mass black
holes, roughly eight orders of magnitude less than the duration of typical
observing sessions.  This gives great statistical power to methods that use
the large number of photons detected in an extended observation.  It becomes
possible to detect variability of the time scale of the mean intervals
between photons, rather than that of Eq.~\ref{t}.

The method proposed here is an adaptation of the method used \citep{K24}
to quantify the dynamics of objects, such as Fast Radio Bursts, Soft Gamma
Repeaters and glitching pulsars, that exhibit repeating outbursts.  That
method fitted log-normal functions to the distributions of intervals
between outbursts.  The width of such a function is a robust measure
(independent of observing threshold if the distribution of outburst
strengths is a power law) of the underlying dynamics, in analogy to the
critical exponents of renormalization group theory \citep{PV02}.

Here we examine and fit log-normals to the distributions of the intervals
between individual detected photons.  For steady sources the statistics are
those of shot noise, but for variable sources there is an excess of longer
intervals during periods of lower intensity and of shorter intervals during
periods of higher intensity; the widths of the log normal fits are wider.
Because the frequency of long intervals is an exponentially decreasing
function of the interval length when the intensity is constant, this method
is sensitive to the presence of periods of lower intensity with
exponentially more long intervals, and therefore to variability.
\section{Synthetic Data and Results}
\subsection{Method}
\begin{figure}
	\centering
	\includegraphics[width=\columnwidth]{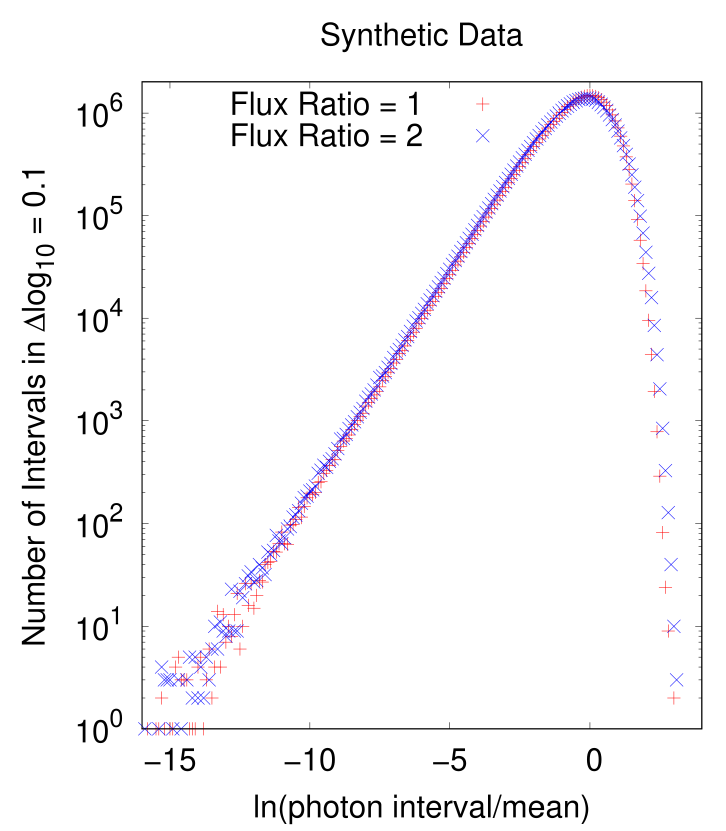}
	\caption{\label{syn-164}The distribution of $4 \times 10^7$
	intervals for constant flux (Flux Ratio = 1) if half the photons
	are detected at a mean rate $\sqrt{2}$ times the overall mean and
	half at a rate $1/\sqrt{2}$ of the overall mean (Flux Ratio = 2).
	Varying the flux broadens the distribution.  This is barely
	visible at small intervals, but large for long intervals because
	of the slower exponential decay of the distribution with
	increasing interval when the mean rate is lower.}
\end{figure}

We use a double precision random number generator to generate synthetic
data.  Fig.~\ref{syn-164} shows the distributions of photon intervals for a
uniform flux (shot noise) with equal numbers of photons detected during
periods whose flux differs by a factor of two (the higher rate occurring
during 1/3 of a nominal observation and the lower rate during 2/3 of that
period).  The distributions are very similar for photon intervals close to
or below the mean, but varying the rate broadens the distribution.

\begin{figure}
	\centering
	\includegraphics[width=\columnwidth]{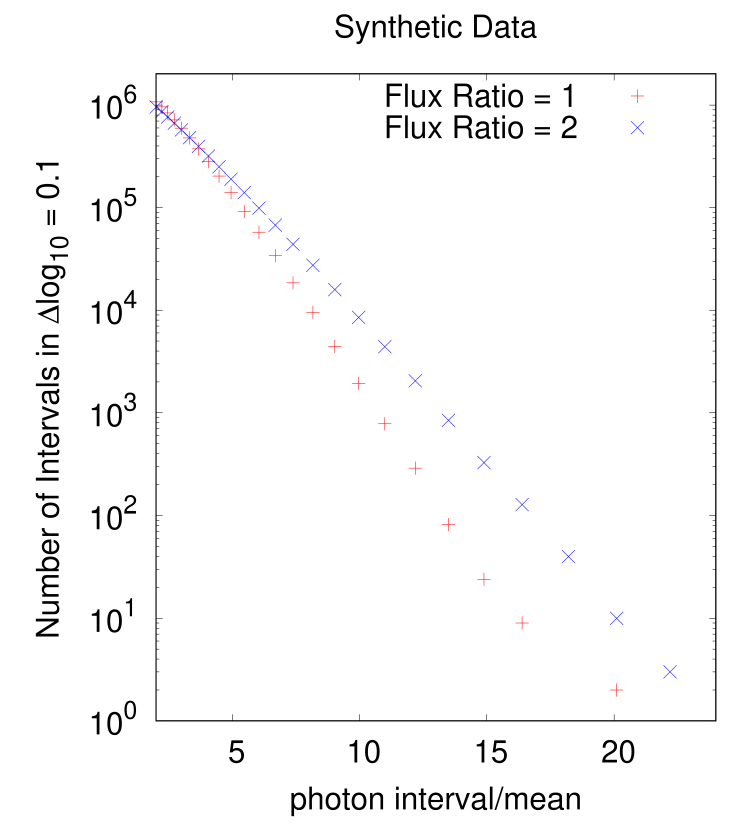}
	\caption{\label{synr}The distribution of longer intervals for
	$4 \times 10^7$ detected photons, half at constant flux (Flux Ratio
	= 1) and half evenly divided between photon detections at rates of
	$\sqrt{2}$ of the mean and $1/\sqrt{2}$ of the mean (Flux Ratio =
	2).  Note the exponential $y$-axis.}
\end{figure}

The difference is large at intervals more than a few times the mean
because of the slower exponential decay of the distribution during the
period of low flux.  This is obvious in Fig.~\ref{synr}, which replots the
synthetic data of Fig.~\ref{syn-164} for the longer intervals.
\subsection{Log-Normal Fits}
\begin{figure}
	\centering
	\includegraphics[width=\columnwidth]{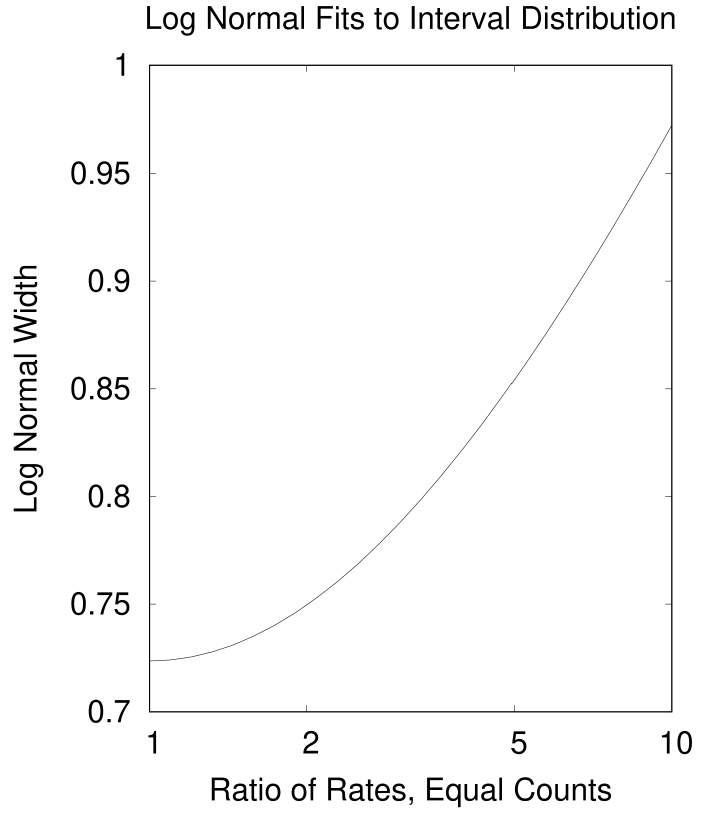}
	\caption{\label{varrate}Width $\sigma$ of log-normal fit to interval
	distribution as a function of the ratio of fluxes in high and low
	flux states; same number of photons detected in each state.}
\end{figure}

The widths of log-normal fits to synthetic data with different ratios of
the high and low rates are shown in Fig.~\ref{varrate}, where again the
numbers of photons detected in the high and low rate periods are equal.
A ratio of 1 corresponds to a constant flux (shot noise).  Larger ratios
correspond to briefer periods of high flux (flares) separated by longer
periods of low flux.

\begin{figure}
	\centering
	\includegraphics[width=\columnwidth]{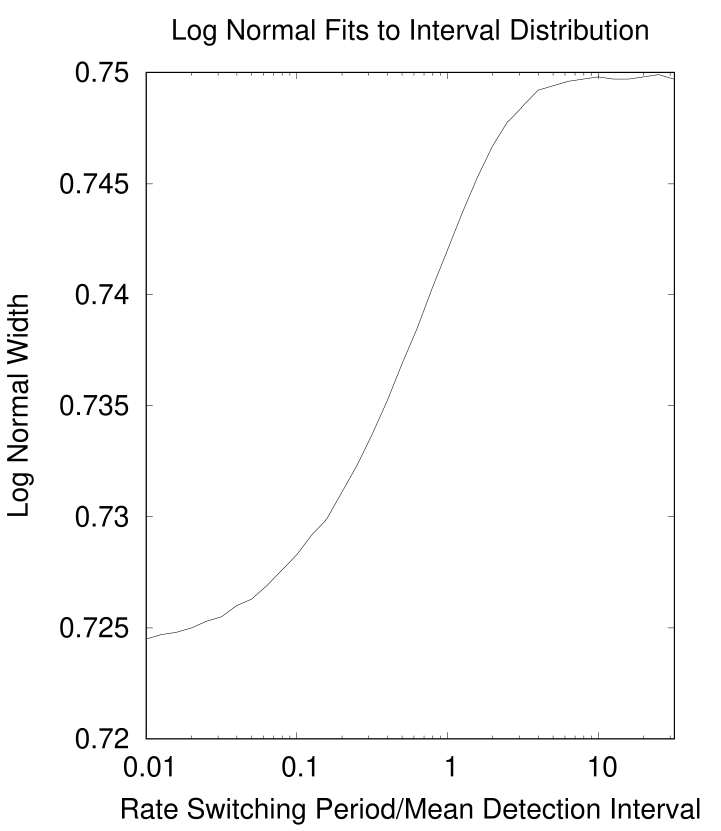}
	\caption{\label{switch} Width of log-normal fit to interval
	distribution when detection rates switch between $\sqrt{2}$ of mean
	and $1./\sqrt{2}$ of mean, with the same number of detections at
	each rate, for a total of $2.5 \times 10^8$ synthetic detected
	photons.}
\end{figure}

We then consider rapid switching between higher and lower flux states.
If the switching occurs on a time $\delta$ short compared to the mean photon
detection interval $\Delta$ the statistics approach those of shot noise
because the effect of switching is to displace the time of detection by an
average $\delta/2 \ll Delta$, which has little effect on the distribution of
intervals ${\cal O}(\Delta) \gg \delta$.  If switching is slow ($\delta \gg
\Delta$) the statistics approach those of Figs.~\ref{syn-164}, \ref{synr} and
\ref{varrate}, in which $2 \times 10^7$ synthetic counts occur in each of
the separate, effectively almost infinitely long, high and low flux states.
Results are shown in Fig.~\ref{switch}.
\subsection{Other Distributions}
The previous examples have assumed equal numbers of detected photons in
high and low flux states.  That may be a useful approximation for sources
that show occasional flares rising from a steadier background.  But there
may be sources nearly all of whose fluence occurs within flares (for
example, Soft Gamma Repeaters).  The width of a log-normal fit to the
interval distribution would be determined by its value during the flares;
the width would not reveal the presence of gaps with negligible, or no,
fluence.

\begin{figure}
	\centering
	\includegraphics[width=\columnwidth]{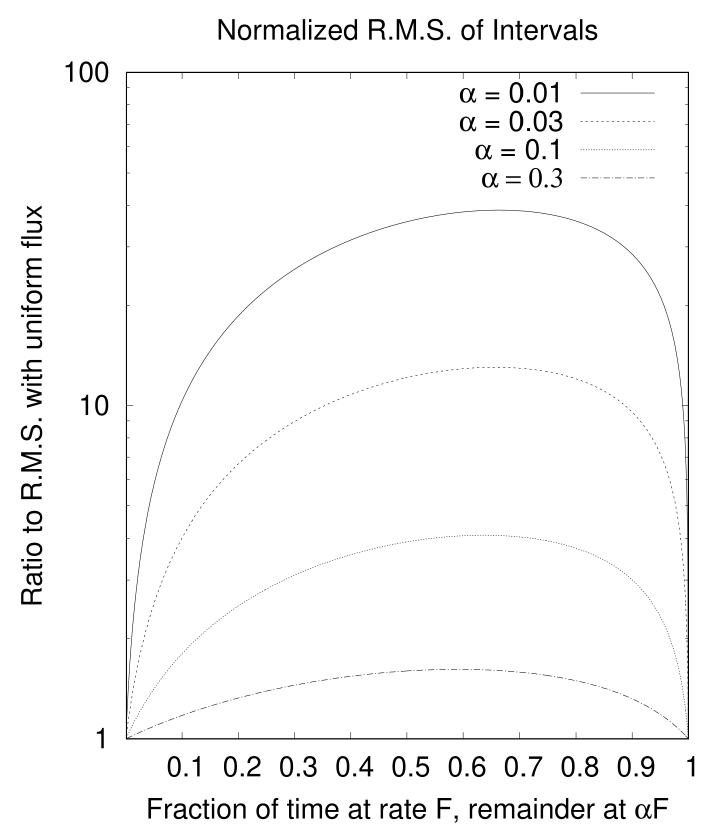}
	\caption{\label{rmsfig}Root-mean-square intervals between detected
	photons, normalized to the mean interval and divided by the value
	for constant flux, as a function of the fraction of time in higher
	flux state.  $\alpha$ is the ratio of the flux in the low state to
	that in the high state, and shot noise statistics are assumed in
	both states.  This measure is sensitive time in the low flux state
	because intervals are, on average, then longer.}
\end{figure}

Other parameters may then be more useful descriptors.  For example, consider
the normalized root-mean-square interval, which for shot noise statistics is
\begin{equation}
\label{rms}
	{\sqrt{\langle (\Delta t)^2 \rangle} \over \langle \Delta t \rangle}
	= \sqrt{2}.
\end{equation}
For a source with mean detection rate $F$ during a fraction $f$ of observing
time, and $\alpha F$ during the remaining fraction $1-f$, both with shot
noise statistics, the normalized root-mean-square interval between
detections, divided by that (Eq.~\ref{rms}) of a steady source,
\begin{equation}
	{\sqrt{\langle (\Delta t)^2 \rangle} \over \langle \Delta t \rangle}
	= \sqrt{2} \sqrt{f + {1-f \over \alpha^2}}
	\left(f + (1-f)\alpha\right).
\end{equation}
This is plotted in Fig.~\ref{rmsfig}.  Unsurprisingly, the root-mean-square
interval is sensitive to periods of low flux when intervals between
detected photons are long.

\begin{figure}
	\centering
	\includegraphics[width=\columnwidth]{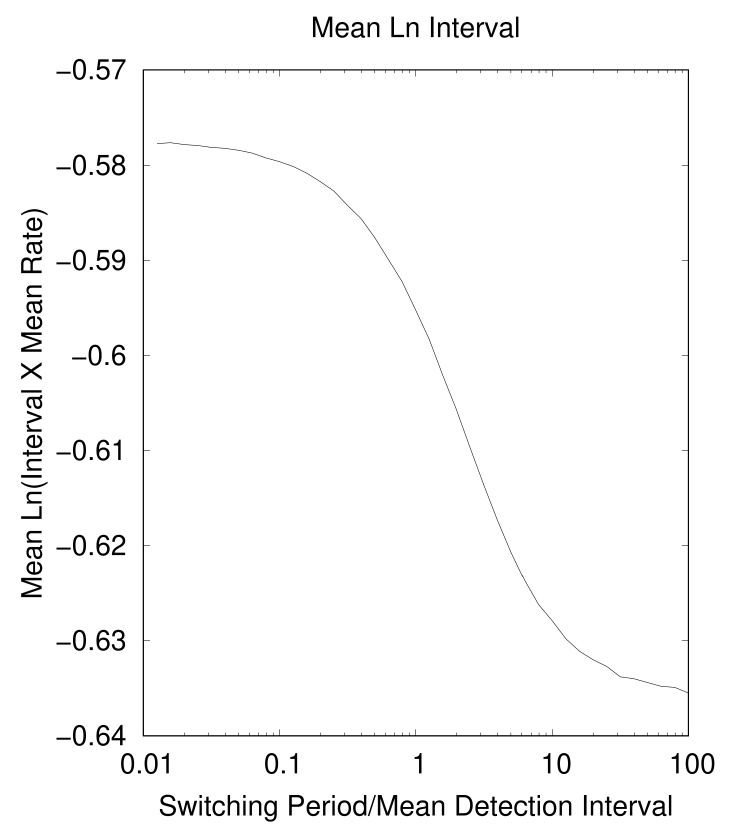}.
	\caption{\label{logs} Mean natural log of the ratio of the interval
	between detections to the mean interval, when detection rates switch
	between $\sqrt{2}$ of mean and $1/\sqrt{2}$ of mean, with the same
	number of detections at each rate, as a function of the switching
	period.  There are a total of $8 \times 10^7$ synthetic detected
	photons for each value of the mean count rate.}
\end{figure}

Other measures are possible.  For example, the average logarithm of the
ratio of the intervals between detected photons to the mean interval
depends on the source's variability.  Fig.~\ref{logs} shows the dependence
of this average on the period of variation between states whose detection
rates differ by a factor of 2, with statistically equal numbers of
detections in each state, the same conditions as in Fig.~\ref{switch}

\section{Example: RXTE Observation of Cyg X-1}
\begin{figure}
	\centering
	\includegraphics[width=\columnwidth]{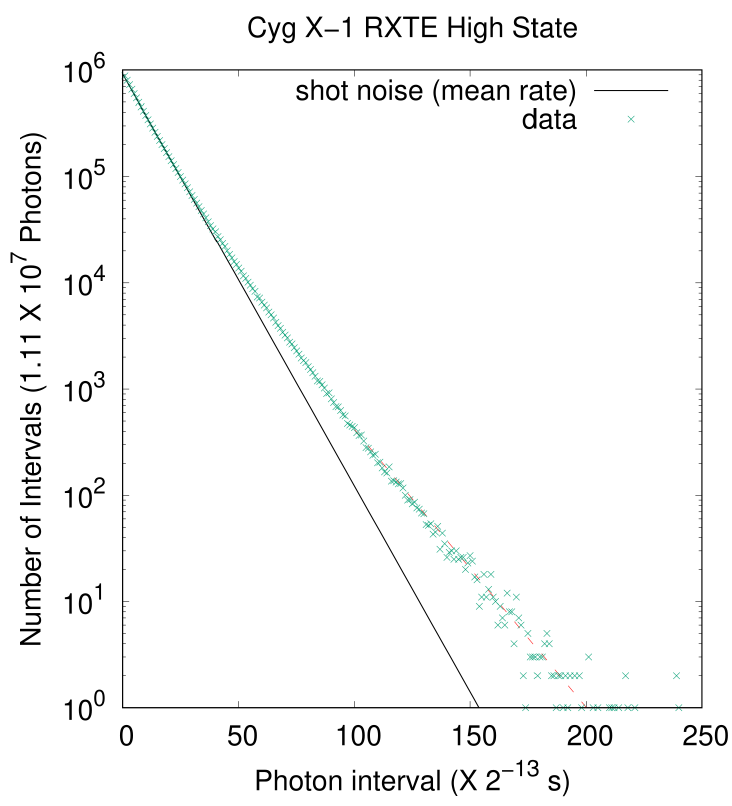}
	\caption{\label{cygx-1}Distribution of intervals between RXTE
	detections \citep{N99} of 11077794 photons from Cyg X-1 during a net
	4.3 hours of observation distributed over 8.1 clock hours, with
	temporal resolution of $2^{-13}\,\text{s} \approx 122\,\mu$s, the
	instrumental time unit.  The solid line is the exponential that
	would describe the distribution of long intervals for shot noise at
	the mean detection rate.  The dashed red line is an approximate 
	exponential fit to the distribution of longer intervals.  Its slope
	indicates that at some times during the observation the flux was
	about 30\% less than its mean, and its intercept with the shot noise
	distribution indicates that about 1\% of the counts occur during
	these times of low flux.}
\end{figure}

As an illustration of the methods developed here, Fig.~\ref{cygx-1} shows
the distribution of intervals between detections of X-rays from Cyg X-1.
The data come from a Rossi X-ray Timing Explorer (RXTE) observation
conducted on 1996 October 22 (Observation ID 10241) while Cyg X-1 was in a
spectrally hard state. The lightcurves were constructed from Proportional
Counter Array channels 36 to 76 (approximately 13--28 keV) using event mode
data where individual events were tagged with a time resolution of $2^{-13}$
s.

This data set and specific energy band were chosen because the integrated
observing time was long, all the detectors on the spacecraft were still
fully operational, the event mode had the highest time resolution for that
observation, and the chosen energy band is strongly dominated by events from
the source and not any background component.  A description of this specific
observation, and the full data modes employed, can be found in \citet{N99},
and the data themselves are available from the High Energy Astrophysics
Archives.

The temporal resolution of $2^{-13} \text{s} \approx 122 \mu$s was coarse
enough that about 10\% of the intervals recorded more than one photon and
the discreteness of the possible intervals precludes accurate log-normal
fits.  However, comparison of the distribution of long intervals to a shot
noise distribution at the mean detection rate shows an excess of long
intervals in the data.  This implies a variable flux, with the excess
occurring during periods of lesser flux; {\it cf.\/} Fig.~\ref{synr}.
Fitting the slope at longer intervals to an exponential suggests a flux
about 30\% less than the mean.  This is not a new conclusion; the
variability of Cyg X-1 has long been known and studied \citep{R71,W78}, but
this analysis demonstrate the power of studying the distribution of photon
detection intervals.
\section{Conclusions}
The development of time-tagging astronomical photon detectors, particularly
the X-ray detectors of RXTE and NICER, enables the measurement of flux
variability on time scales as short as the reciprocal of the photon count
rate.  This information is encoded in the distribution of intervals between
successive detected photons.

It is not required to minimize statistical fluctuations by detecting many
photons on the time scale of the variability.  It is only necessary that
the photons be time-tagged with a precision better than the variability time
scale.  There is no magic here; just as many photons must be detected, but
they may be detected cumulatively over a longer time.  Rather than requiring
$(S/N)^2/f^2$ detections per variability time scale, only about one is
required.  The price paid is that, unlike direct measurement of the flux,
the method does not produce a ``light curve''; the presence of variability
may be detected, but not the dependence of flux on time.
\section*{Acknowledgements}
This research has made use of data and/or software provided by the High
Energy Astrophysics Science Archive Research Center (HEASARC), which is a
service of the Astrophysics Science Division at NASA/GSFC.

We thank T. Piran for useful discussions.
\section*{Data Availability}
This theoretical study produced no new data.  The codes used to calculate
the figures will be provided on request.



\bsp	
\label{lastpage}
\end{document}